\documentclass[a4paper,12pt]{article}

\usepackage{graphicx}
\usepackage{amsmath}
\usepackage{multirow}
\usepackage{float} 

\begin{document}

\begin{flushright}
preprint SHEP-10-44\\
\today
\end{flushright}
\vspace*{1.0truecm}

\begin{center}
{\large\bf Muon Signals of Very Light CP-odd Higgs states\\[0.15cm]
of the NMSSM at the LHC}\\
\vspace*{1.0truecm}
{\large M. M. Almarashi and S. Moretti}\\
\vspace*{0.5truecm}
{\it School of Physics \& Astronomy, \\
 University of Southampton, Southampton, SO17 1BJ, UK}
\end{center}

\vspace*{2.0truecm}
\begin{center}
\begin{abstract}
\noindent 
We study here the $\mu^+\mu^-$ decay mode of a very light CP-odd Higgs boson of the NMSSM, $a_1$,
produced in association with a bottom-antibottom pair and find that, despite small event rates, 
a significant signal should be extractable from the SM background at the LHC with high luminosity. 
\end{abstract}
\end{center}

\newcommand{\nn}{\nonumber}
\section{Introduction}
\label{sect:intro}

In the Next-to-Minimal Supersymmetric Standard Model (NMSSM) \cite{review}, as 
a result of the introduction of an extra complex singlet scalar
field, which only couples to the two MSSM-type Higgs doublets, the ensuing
Higgs sector comprises a total of seven mass
eigenstates: a charged pair $h^\pm$, three CP-even Higgses
$h_{1,2,3}$ ($m_{h_1}<m_{h_2}<m_{h_3}$) and two CP-odd
Higgses $a_{1,2}$ ($m_{a_1}<m_{a_2}$). Consequently, Higgs
phenomenology in the NMSSM is plausibly different from that of the
MSSM, i.e, the minimal realisation of Supersymmetry (SUSY). 

In particular, over the past few years, there have been attempts
to extend the so-called `No-lose theorem' of the MSSM -- stating 
 that at least one MSSM Higgs boson should be observed via
the usual SM-like production and decay channels at the Large Hadron Collider (LHC) 
 throughout the entire MSSM parameter space \cite{NoLoseMSSM} --
to the case of the NMSSM \cite{NMSSM-Points,NoLoseNMSSM1,Shobig2}. From this perspective,
it was realised that at least one NMSSM Higgs boson should remain observable 
at the LHC over the NMSSM parameter space that does not allow any Higgs-to-Higgs 
decay. However, when the only light non-singlet (and, therefore, potentially visible) CP-even
Higgs boson, $h_1$ or $h_2$, decays mainly to two very light 
CP-odd Higgs bosons, $h_{1,2}\to a_1 a_1$, one
may not have a Higgs signal of statistical significance at the LHC \cite{dirk}. In fact, 
further violations to the theorem may well occur if
one enables Higgs-to-SUSY particle decays (e.g., into neutralino pairs, yielding invisible Higgs signals)
\cite{Cyril,NMSSM-Benchmarks}.

While there is no conclusive evidence on whether a `No-lose theorem' can be proved for the NMSSM
at the LHC, there has also been put forward an orthogonal approach: to see when a, so to say,
`More-to-gain theorem' for the LHC \cite{Shobig1,Erice,CPNSH,Almarashi:2010jm} can be formulated 
within the NMSSM. That is, whether there exist regions
of the NMSSM parameter space where more and/or different Higgs
states of the NMSSM are visible at the LHC than those available within the MSSM.
 
In our attempt to overview both such possibilities, we consider here the case
of the $\mu^+\mu^-$  decay channel of a light neutral CP-odd Higgs boson produced in 
association with $b$-quark pairs at the LHC. This work complements the one carried out 
in \cite{Almarashi:2010jm}, which concentrated on $a_1\to \tau^+\tau^-$ and $\gamma\gamma$ decays.

The plan of this paper is as follows. In Sect.~\ref{sect:scan} we describe the parameter space scans
performed. In Sect.~\ref{sect:rates} we discuss inclusive event rates.
In Sect.~\ref{sect:S2B} we describe the signal-to-background analysis performed and introduce
some benchmark points where an $a_1$ signal can be extracted
via the $\mu^+\mu^-$ decay mode. Finally, in Sect.~\ref{sect:summa}, we summarise and
conclude.  

\section{\large Parameter Space Scan}
\label{sect:scan}

For our study of the NMSSM Higgs sector
we have used NMSSMTools \cite{NMHDECAY,NMSSMTools}. This package
computes the masses, couplings and decay widths into two particle final states of all the Higgs
bosons of the NMSSM in terms of its input parameters, which can be taken at either the Electro-Weak (EW)
or grand unification scale. NMSSMTools also takes into account theoretical consistency as well as
experimental limits from negative Higgs searches at LEP \cite{LEP} and Tevatron\footnote{Speculations
of an excess at LEP which could be attributed to NMSSM Higgs bosons are found in \cite{excess}. Very light 
CP-odd Higgs bosons of the NMSSM could also be produced in some rare hadron decays \cite{rare}},
including the unconventional channels relevant for the NMSSM only. 

Here, instead of postulating unification or taking into account the 
SUSY breaking mechanism, we fixed the soft SUSY breaking terms to a very high value, 
so that they give a small or no contribution at all to the outputs of the parameter scans. 
Consequently, we are left with six free parameters at the EW scale, uniquely defining the NMSSM Higgs sector
at tree-level. Our  parameter space is in particular defined through the Yukawa couplings $\lambda$ and
$\kappa$, the soft trilinear terms $A_\lambda$ and $A_\kappa$, plus 
tan$\beta$ (the ratio of the 
Vacuum Expectation Values (VEVs) of the two Higgs doublets) and $\mu_{\rm eff} = \lambda\langle S\rangle$
(where $\langle S\rangle$ is the VEV of the Higgs singlet). 

In order to make a comprehensive study of the NMSSM parameter space,
we have used the NMHDECAY code to scan over the aforementioned six parameters taken in the following 
intervals:
\begin{center}
$\lambda$ : 0.0001 -- 0.7,\phantom{aa} $\kappa$ : 0 --
0.65,\phantom{aa} $\tan\beta$ : 1.6 -- 54,\\ $\mu$ : 100 -- 1000 GeV,\phantom{aa} $A_{\lambda}$ : $-$1000 -- +1000 GeV,\phantom{aa} $A_{\kappa}$ :$-$10 -- 0.\\
\end{center}
Soft terms which are fixed in the scan include:\\
$\bullet\phantom{a}m_{Q_3} = m_{U_3} = m_{D_3} = m_{L_3} = m_{E_3} = 1$ TeV, \\
$\bullet\phantom{a}A_{U_3} = A_{D_3} = A_{E_3} = 1.2$ TeV,\\
$\bullet\phantom{a}m_Q = m_U = m_D = m_L = m_E = 1$ TeV,\\
$\bullet\phantom{a} M_1 = M_2 = M_3 = 1.5$ TeV.\\

In line with the assumptions made in \cite{NMSSM-Points,NoLoseNMSSM1}, the allowed decay 
modes for the CP-odd neutral NMSSM Higgs boson $a_1$ are:
\begin{equation}
a_1\rightarrow \mu^+\mu^-,\tau^+\tau^-,gg,
s\bar s,c\bar c,b\bar b, t\bar t, 
\gamma\gamma,Z\gamma,~{\rm sparticles}.
\end{equation}

We have performed our scan over 10 millions of randomly
selected points in the specified parameter space. The output, as
stated earlier, contains masses, Branching Ratios (BRs) and couplings of
the NMSSM Higgses, for all the points which have passed 
the various experimental and theoretical constraints. The points which violate the latter
are automatically eliminated by NMSSMTools. 

\section{\large Inclusive Event Rates}
\label{sect:rates}

The surviving data points are then used to determine the
cross-sections for NMSSM Higgs hadro-production by using CalcHEP \cite{CalcHEP}
and MadGraph \cite{MadGraph}\footnote{We adopt herein
CTEQ6L \cite{cteq} as parton distribution functions, with scale $Q=\sqrt{\hat{s}}$, the centre-of-mass energy
at parton level, for all processes computed.}, wherein some new modules 
have been implemented for this purpose \cite{Almarashi}. As the SUSY mass scales 
have been arbitrarily set well above the EW one (see above), 
the NMSSM Higgs production modes
exploitable in simulations at the LHC are those involving couplings to
heavy ordinary matter only. Amongst the production channels onset by the latter, 
we focus here on 
\begin{equation}\label{prod}
q\bar q,  gg\to b\bar b~{a_1},
\end{equation}
i.e., Higgs production in association with a $b$-quark pair.

As an initial step towards the analysis of the data, we have computed the production cross-section
times the decay BR  against each of
the six parameters of the NMSSM, as intimated.
We started by computing total (i.e., fully inclusive) 
rates. Figs. 1 and 2 present the results of our scan, the first series of plots illustrating the distribution
of event rates over the six independent NMSSM parameters, $m_{a_1}$ and as a function of the BR of the 
corresponding channel with the second plot displaying the correlations between the $a_1$ 
mass and the di-muon decay rate. It is clear
that the large $\tan\beta$ and small $\mu_{\rm eff}$ (and, to some extent, also small $\lambda$) region is the one most 
compatible with current theoretical and experimental constraints, while the distributions in $\kappa$,
$A_\lambda$ and $A_\kappa$ are rather uniform (top six panes in Fig. 1). From a close look at Fig. 2 it is further clear 
that the BR$(a_1\to\mu^+\mu^-)$ can be of ${\cal O}(10\%)$(${\cal O}(1\%)$)[${\cal O}(0.1\%)$ or less] when 
$2m_\mu<m_{a_1}<2m_\tau$($2m_\tau<m_{a_1}<2m_b$)[$2m_b<m_{a_1}$], respectively. The first region of parameter space 
($m_{a_1}<2m_\tau$) is rather small, the second one
($2m_\tau<m_{a_1}<2m_b$) more significant, yet the widest one is the third one ($2m_b<m_{a_1}$). However, by looking  at the
two bottom panes of Fig. 1, it is remarkable to notice that the event rates are sizable in all such mass regions, topping the
$10^4$ fb level in the two lower mass intervals and the $10^3$ fb level in the higher mass range. Finally, notice that the mass
region below the $\mu^+\mu^-$ threshold is severly constrained \cite{Lebedev}.  

\section{Signal-to-Background Analysis}
\label{sect:S2B}

We perform here a partonic signal-to-background ($S/B$) analysis. We assume $\sqrt s=14$ TeV throughout for the 
LHC energy. Also, in our numerical 
analyses, we have taken $m_b(m_b)=4.214$ GeV and $m_t^{\rm pole}=171.4$ GeV for the (running) bottom- and 
(pole) top-quark mass, 
respectively, while we have input $m_\tau^{\rm pole}=1.777$ GeV and $m_\mu^{\rm pole}=0.1057$ GeV for the (pole) tau- and 
(pole) muon-lepton mass, 
respectively.
After implementing the following 
standard cuts
$$\Delta R (b,\bar b),\Delta R(b,\mu^{+}), \Delta R(\bar b,\mu^{+}),\Delta R(b,\mu^{-}), \Delta R(\bar b,\mu^{-}), \Delta R(\mu^{+},\mu^{-})> 0.4$$
$$\arrowvert\eta(b)\arrowvert,\arrowvert\eta(\bar b)\arrowvert,\arrowvert\eta(\mu^{+})\arrowvert,\arrowvert\eta(\mu^{-})\arrowvert <2.5$$
\begin{equation}
P_{T}(b),P_{T}(\bar b)>20~{\rm{GeV}}, P_{T}(\mu^{+}),P_{T}(\mu^{-})>5~{\rm{GeV}},
\label{cuts:MM}
\end{equation}
we obtain the invariant masses of the $\mu^+\mu^-$ system depicted in Figs. 3--7, where we show the combined
yield of the signal induced by $q\bar q,gg\to b\bar b \mu^+\mu^-$ (via
$g$ and $a_1$ exchange) and of the irreducible background due to $q\bar q,gg\to b\bar b \mu^+\mu^-$ (via
$g$, $\gamma$ and $Z$ exchange), including their interference.
We also show in Figs.~3--7 the top-antitop reducible background,
i.e., $q\bar q,gg\to t\bar t\to b\bar b W^+W^-\to b\bar b \mu^+\mu^- P_T^{\rm{miss}}$.

We 
notice the dominance of the $\gamma\to\mu^+\mu^-$ tail of the irreducible background 
at very small di-muon invariant masses. Furthermore,  the 
reducible background starts reaching its maximum at around $M_W/2$. Finally, the $Z\to\mu^+\mu^-$ peak of the 
irreducible background becomes overwhelming already
at 60 GeV or so. Overall, the $\mu^+\mu^-$ signal yield in such a mass region is of order 75 (for $m_{a_1}$ 
reaching 60 GeV or so) to 1440 (for $m_{a_1}$ starting at 10 GeV or so) signal events over a sizably smaller 
background (notice the logarithmic scale of the
plots and notice that we are assuming, e.g., 300 fb$^{-1}$ of accumulated luminosity). 

While the number of events is not very large, there is potential scope to extract a significant signal over the above interval thanks to the high mass resolution 
that can be achieved using muon pairs. Assuming $\mu^+\mu^-$ resolutions of 1 GeV \cite{reso}, we obtain as signal significances $S/\sqrt B$ (where $S$ and $B$ are
the signal and background rates, respectively, after a given luminosity), as a function of such
a luminosity, those depicted in Fig. 8 (left hand side). The corresponding signal event rates are found instead in the right hand side of the
same figure. From these last results, it is clear that detection at the LHC could occur for $a_1$ masses between 
$\approx10$ and $\approx40$ GeV with rather modest 
luminosity, 30 fb$^{-1}$ or so, while $\approx50(\approx60)$ GeV masses require some 200(300) fb$^{-1}$ while 
heavier states will not be resolvable even at the end
of the LHC era. 

\section{Conclusions}
\label{sect:summa}
Due to introducing a complex singlet superfield, the NMSSM can have a CP-odd Higgs boson with very low mass,
$a_1$.
We have proven that there exist sizable regions of the NMSSM parameter space where this kind of Higgs state, 
with a mixed singlet and doublet nature, could potentially be detected at the LHC if $10$ GeV $<m_{a_1}\le 60$ GeV in the 
$a_1\to\mu^+\mu^-$ mode if the CP-odd Higgs state is produced in association with a $b\bar b$ pair for large 
tan$\beta$ plus small $\mu_{\rm eff}$ and $\lambda$. After a realistic $S/B$ analysis at parton level, 
we have in fact produced results showing that
the extraction of light mass $a_1\to\mu^+\mu$ resonances above both the irreducible and (dominant) reducible background should be feasible using standard
reconstruction techniques \cite{ATLAS-TDR,CMS-TDR}. While more refined
analyses, incorporating parton shower, hadronisation and detector effects, are needed in order to delineate the 
true discovery potential of the LHC
over the actual NMSSM parameter space, we are confident that our results are a step in the right direction to both:
(i) prove the existence of a `More-to-gain theorem' at the CERN collider for the NMSSM with respect to the MSSM
(as $\mu^+\mu^-$ signals from such light Higgs bosons are not at all possible in the latter scenario) and (ii) 
to establish a `No-lose theorem' for the NMSSM at the LHC (as some of the parameter regions where the aforementioned signal can be detected overlap 
with those where  $h_{1,2}\to a_1a_1$ decays might be ineffective in extracting an $h_1$ signal).

Notice that we have explored here the three mass regimes $2m_\mu < m_{a_1} < 2m_\tau$, 
$2m_\tau < m_{a_1}<2m_b$ and $2m_b<m_{a_1}$. The first interval in unresolvable because of a large irreducible 
background due to soft photons splitting into $\mu^+\mu^-$ pairs. The second one can actually be resolved only
just below its upper end, i.e., close to the $2m_b$ threshold, above which $a_1$ state remains detectable for
several tens of GeV. However, when $m_{a_1}>60$ GeV, it is the combination of both the irreducible and reducible 
background induced by on-shell $Z$ bosons that prevents detection. 
 
\section*{Acknowledgments}
This work is 
supported in part by the NExT Institute. M. M. A. acknowledges
a scholarship granted to him by Taibah University (Saudi Arabia).

\newpage

\begin{figure}[H]
 \centering\begin{tabular}{cc}
 \includegraphics[scale=0.58]{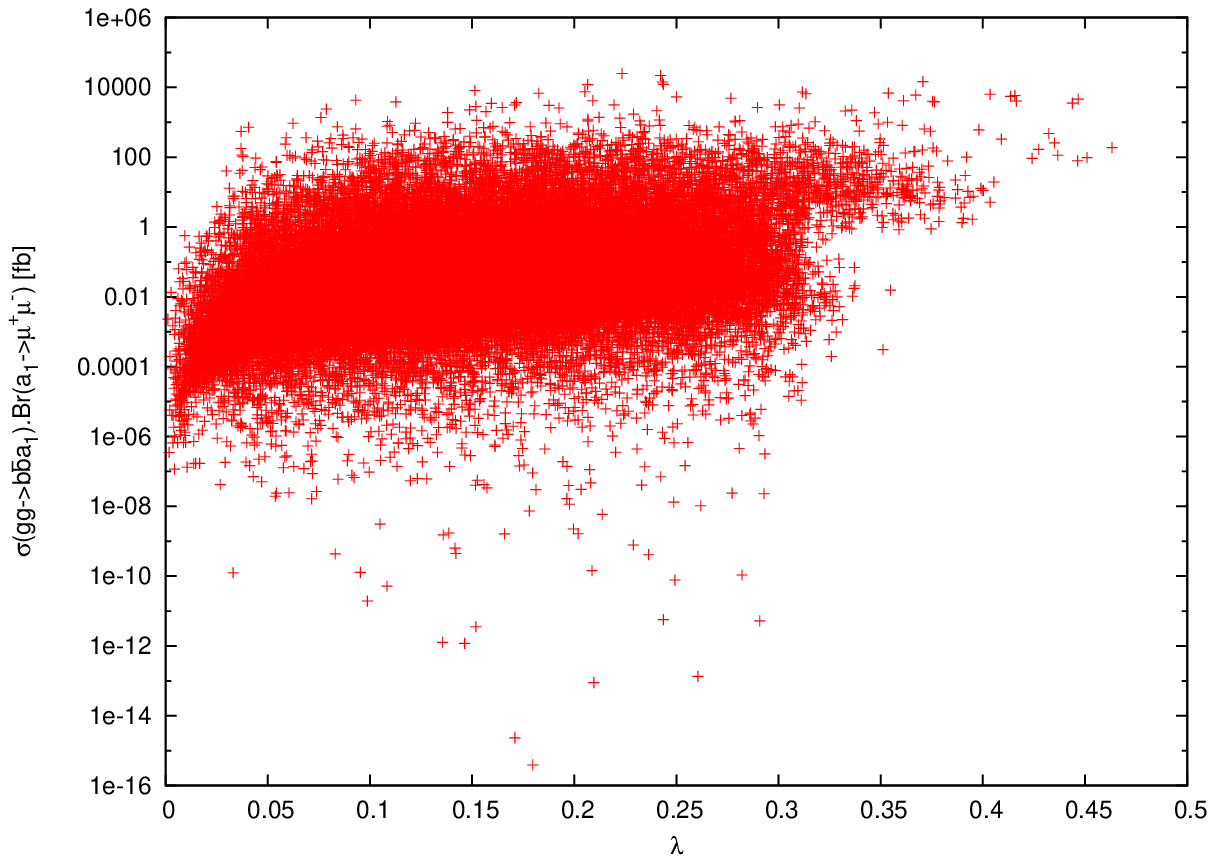}&\includegraphics[scale=0.58]{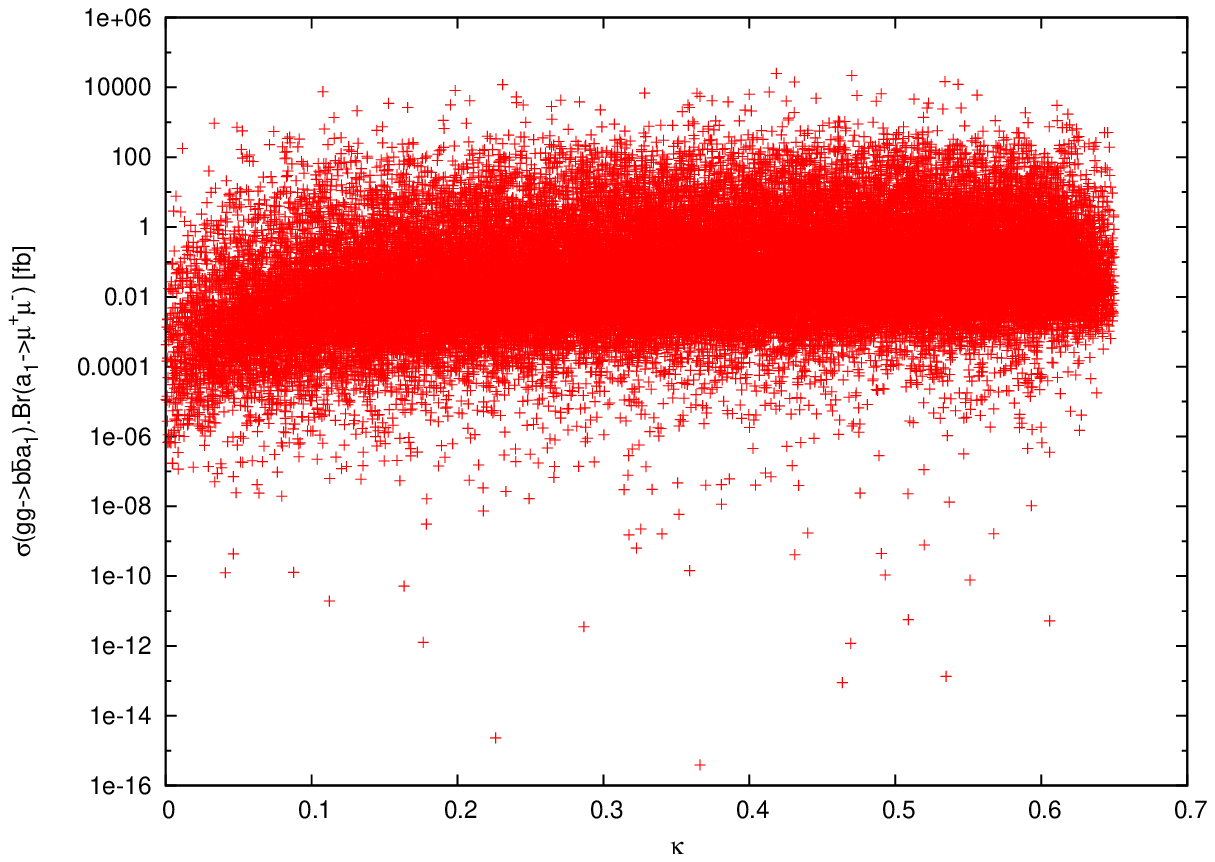}\\
 \includegraphics[scale=0.58]{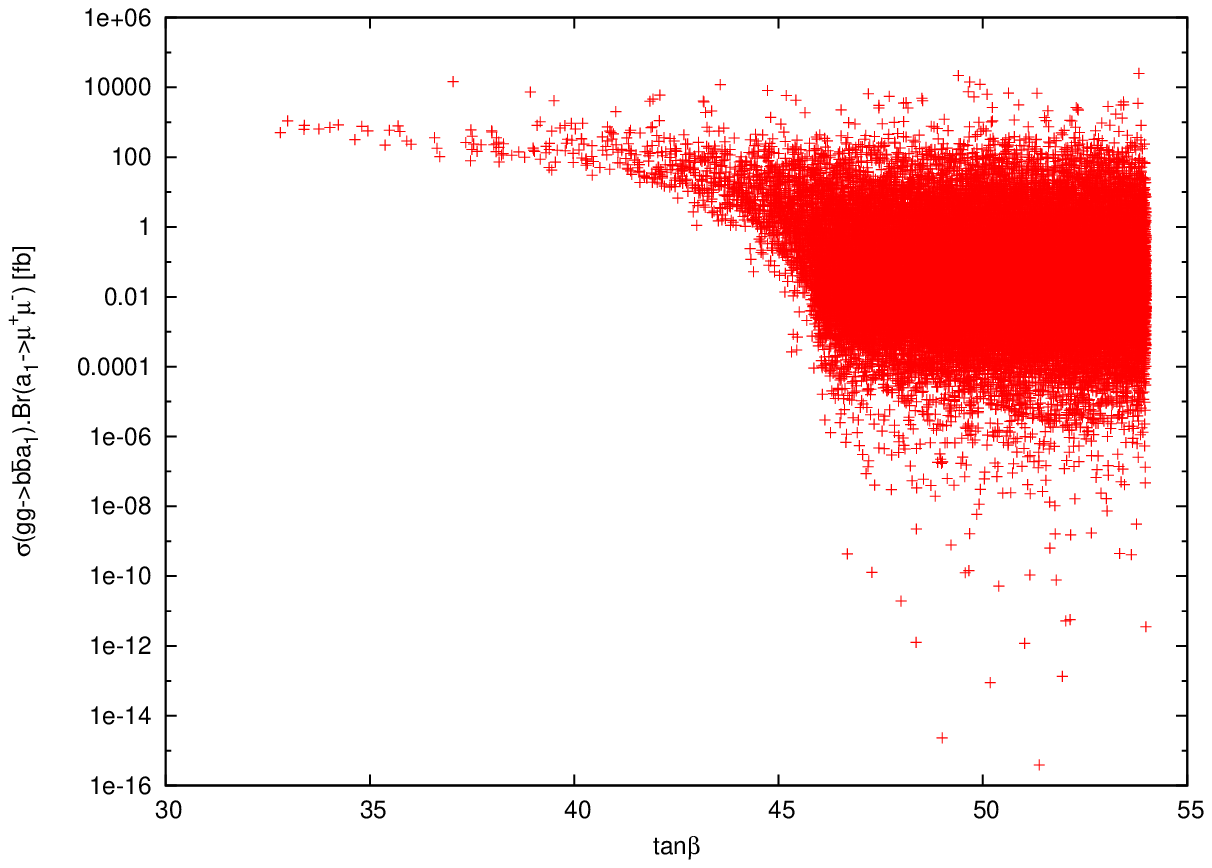}&\includegraphics[scale=0.58]{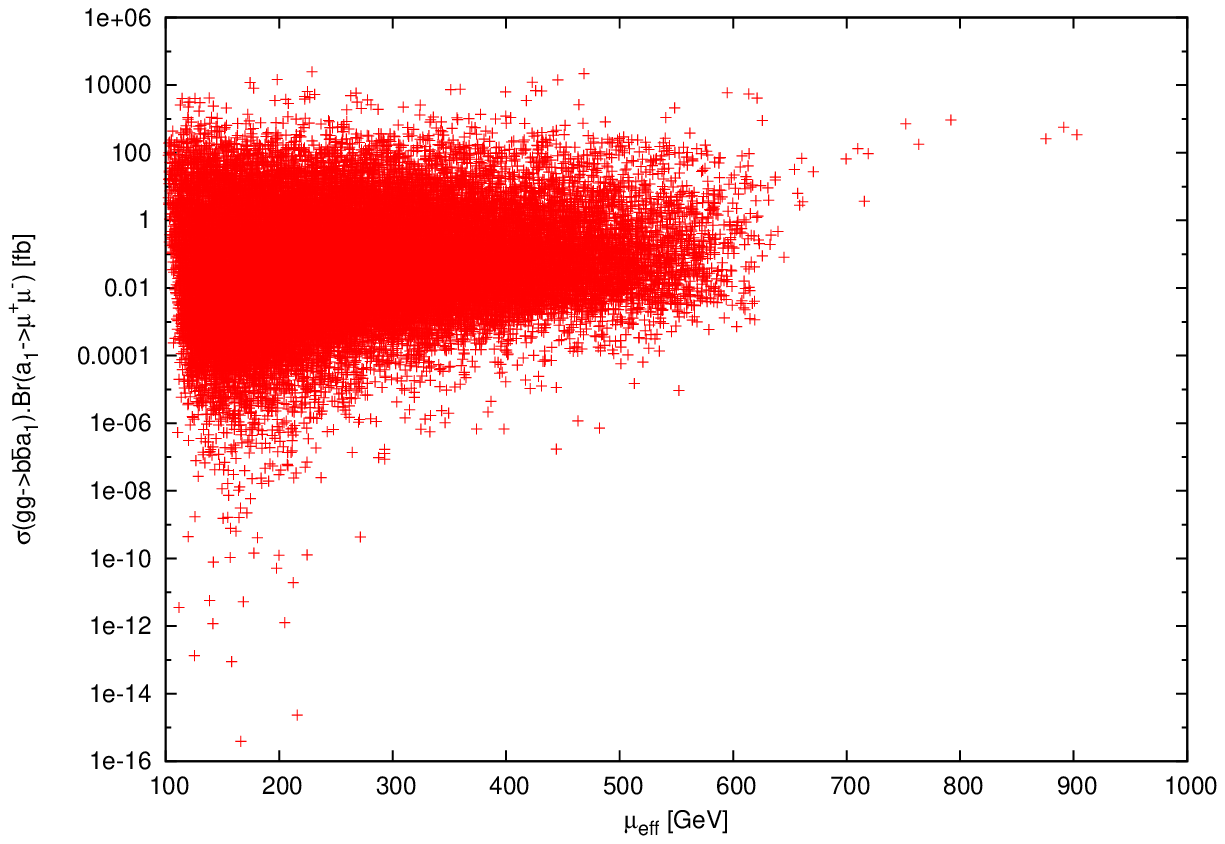}\\
 \includegraphics[scale=0.58]{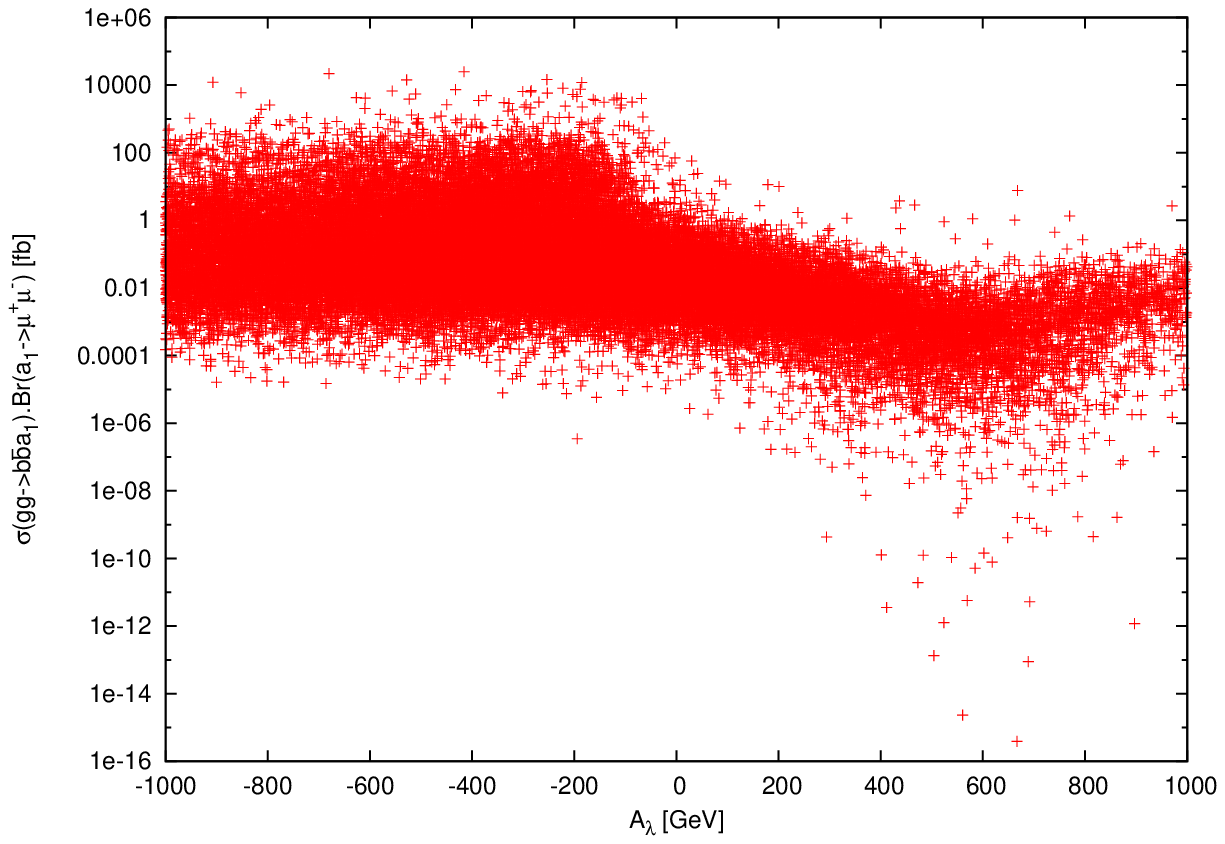}&\includegraphics[scale=0.58]{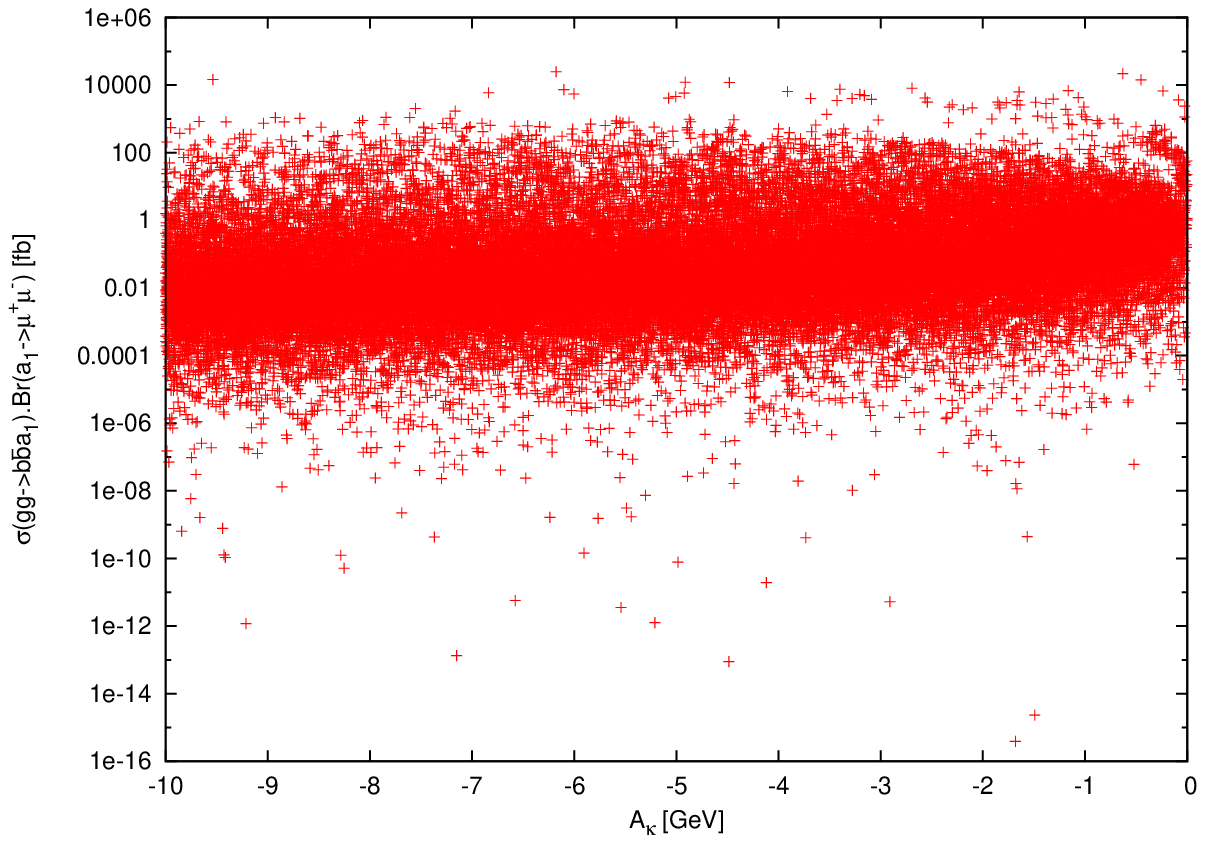}\\
 \includegraphics[scale=0.58]{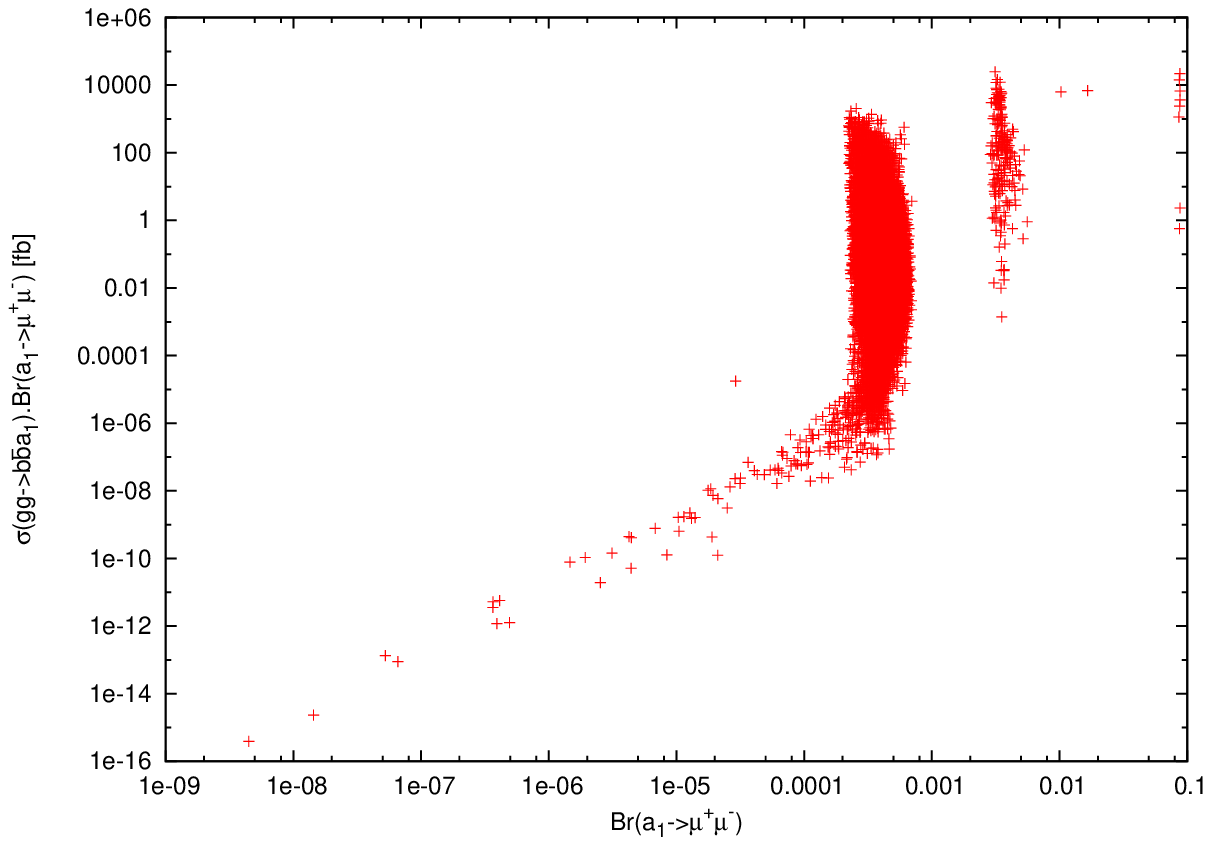}&\includegraphics[scale=0.58]{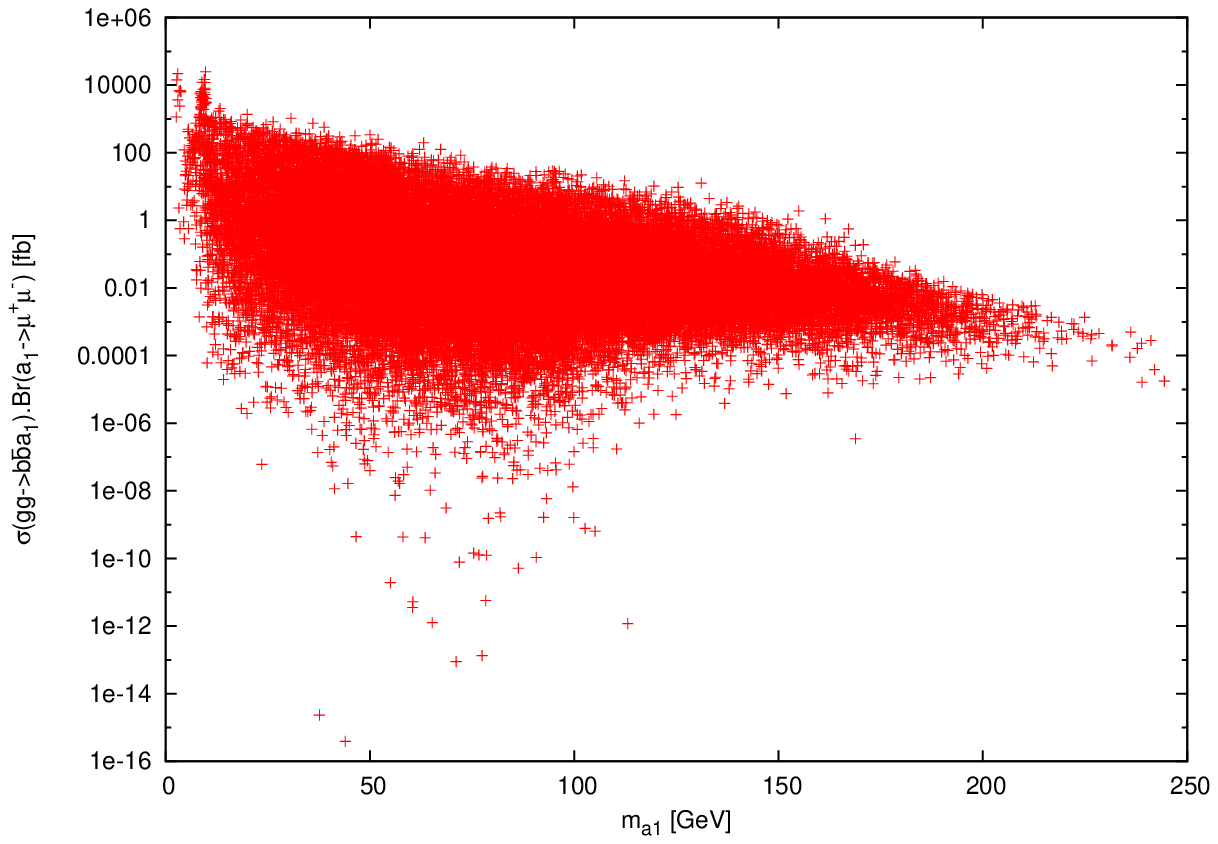}
 \end{tabular}
\label{fig:sigma-scan2muon}
\caption{The rates for $\sigma(q\bar q,gg\to b\bar b {a_1})~{\rm BR}(a_1\to \mu^+\mu^-)$ as a function of $\lambda$, 
$\kappa$, $\tan\beta$, $\mu_{\rm eff}$, $A_\lambda$, $A_\kappa$, BR$(a_1\to \mu^+\mu^-)$ and $m_{a_1}$. }
\end{figure}

\begin{figure}
 \centering\begin{tabular}{c}
  \includegraphics[scale=0.55]{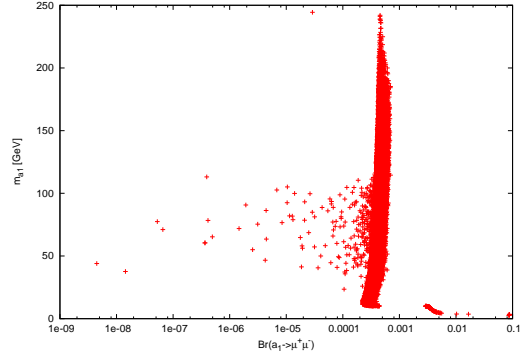}
 \end{tabular}
\label{fig:mass-scan2mu}
\caption{The CP-odd Higgs mass $m_{a_1}$ as a function of BR$(a_1\to \mu^+\mu^-)$.}
\end{figure}

\begin{figure}
 \centering\begin{tabular}{c}

 \includegraphics[scale=0.55]{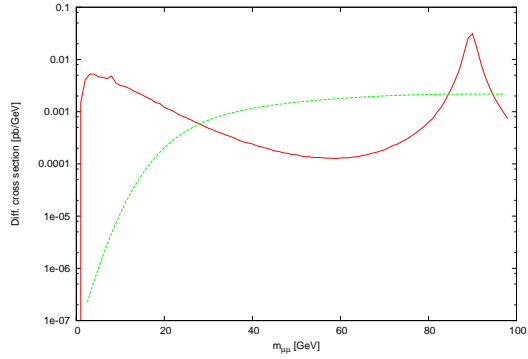}
  \end{tabular} 
\label{fig:MM2}  
\caption{The differential cross section in the $\mu^+\mu^-$ channel for $m_{a_1}$=9.76 GeV as a function of $m_{\mu\mu}$, after the cuts in (\ref{cuts:MM}) 
 with $\lambda$ = 0.22341068, $\kappa$ = 0.4184933, tan$\beta$ = 53.819484, $\mu$ = 228.94259,
 $A_\lambda$ = -415.57365 and $A_\kappa$ = -6.1773643. 
The solid line represents the signal and irreducible background together whereas the dashed line is the 
$t\bar t$ background. (Notice that here $2m_b^{\rm pole}>m_{a_1}$.)} 
 
\end{figure}

\begin{figure}
 \centering\begin{tabular}{c}

 \includegraphics[scale=0.55]{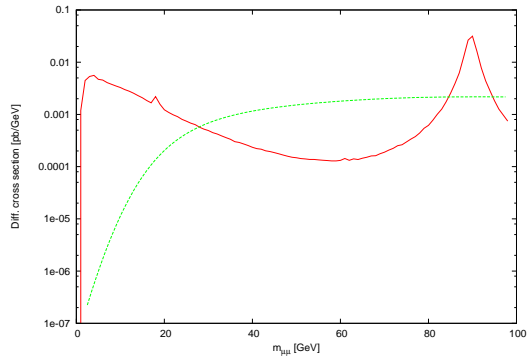}
 \end{tabular}  
\label{fig:MM3}
\caption{The differential cross section in the $\mu^+\mu^-$ channel for $m_{a_1}$=19.98 GeV as a function of  $m_{\mu\mu}$, after the cuts in (\ref{cuts:MM}) 
 with $\lambda$ = 0.075946278, $\kappa$ = 0.11543578, tan$\beta$ = 51.507125, $\mu$ = 377.4387,
 $A_\lambda$ = -579.63592 and $A_\kappa$ = -3.5282881. The solid line represents the signal and irreducible background together whereas the dashed line is the $t\bar t$ background.}
  
\end{figure}

\begin{figure}
 \centering\begin{tabular}{c}
 \includegraphics[scale=0.60]{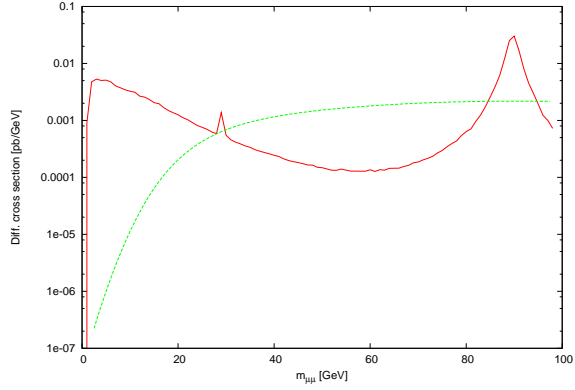}
   \end{tabular} 
\label{fig:MM4}
  \caption{The differential cross section in the $\mu^+\mu^-$ channel for $m_{a_1}$=30.67 GeV as a function of  $m_{\mu\mu}$, after the cuts in (\ref{cuts:MM}) 
 with $\lambda$ = 0.10861169, $\kappa$ = 0.4654168, tan$\beta$ = 48.063727, $\mu$ = 222.99377,
 $A_\lambda$ = -952.59787 and $A_\kappa$ = -7.2147327. The solid line represents the signal and irreducible background together whereas the dashed line is $t\bar t$ background.}
 
\end{figure}

\begin{figure}
 \centering\begin{tabular}{c}
            
 \includegraphics[scale=0.60]{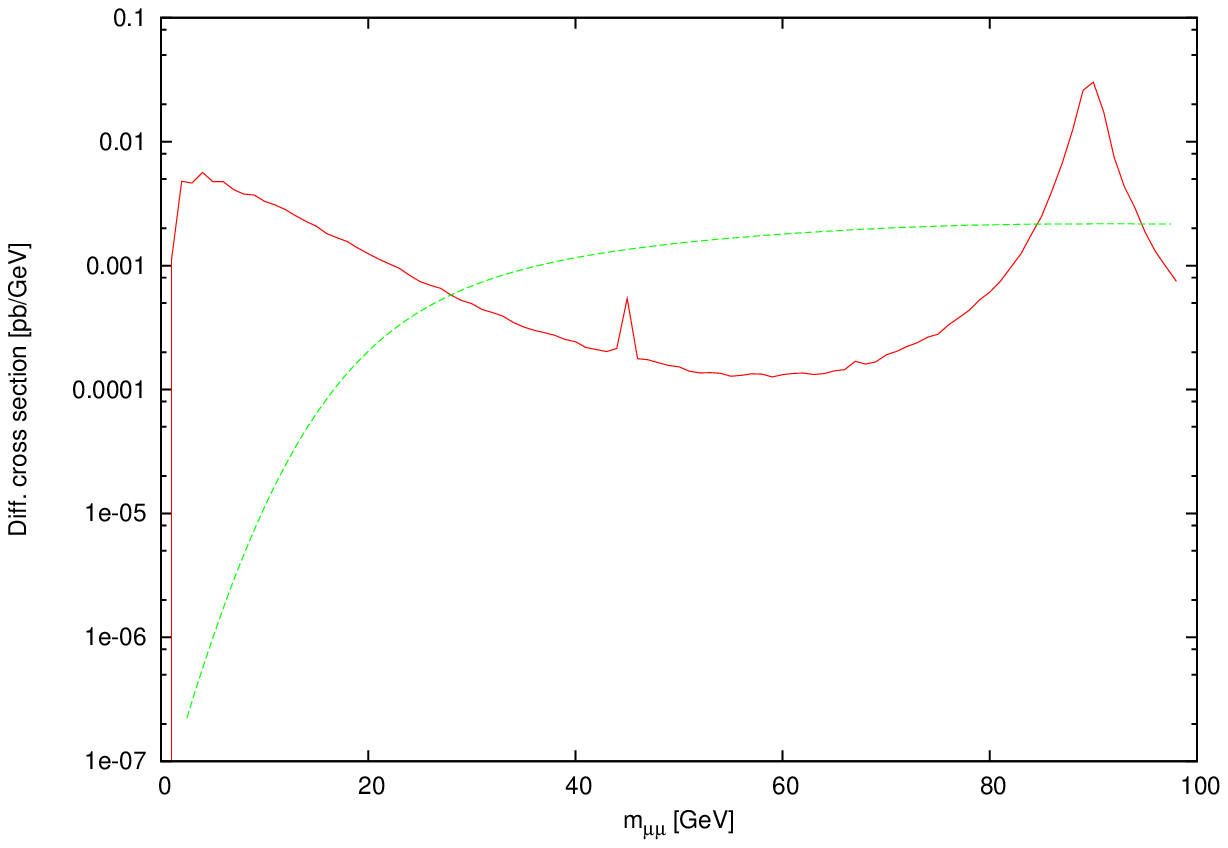}
  \end{tabular}
\label{fig:MM5}  
 \caption{The differential cross section in the $\mu^+\mu^-$ channel for $m_{a_1}$=46.35 GeV as a function of $m_{\mu\mu}$, after the cuts in (\ref{cuts:MM}) 
with $\lambda$ = 0.14088263, $\kappa$ = 0.25219468, tan$\beta$ = 50.558484, $\mu$ = 317.07532,
 $A_\lambda$ = -569.60665 and $A_\kappa$ = -8.6099538. The solid line represents the signal and irreducible background together whereas the dashed line is the $t\bar t$ background.}

\end{figure}

\begin{figure}
 \centering\begin{tabular}{c}
 \includegraphics[scale=0.60]{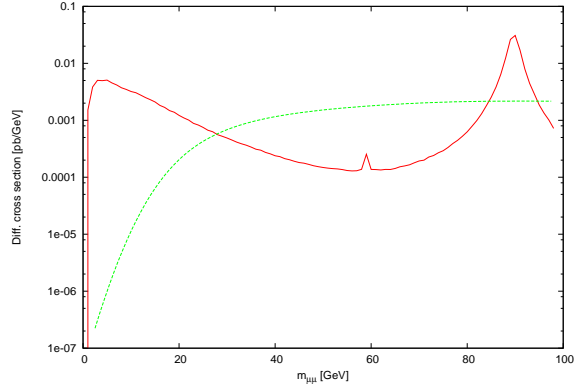}
   \end{tabular}
\label{fig:MM6} 
 \caption{The differential cross section in the $\mu^+\mu^-$ channel for $m_{a_1}$=60.51 GeV as a function of $m_{\mu\mu}$, after the cuts in (\ref{cuts:MM}) 
with $\lambda$ = 0.17410656, $\kappa$ = 0.47848034, tan$\beta$ = 52.385408, $\mu$ = 169.83139,
 $A_\lambda$ = -455.85097 and $A_\kappa$ = -9.0278415. The solid line represents the signal and irreducible background together whereas the dashed line is the $t\bar t$ background.}
  
\end{figure}

\begin{figure}
 \centering\begin{tabular}{cc}
    \includegraphics[scale=0.60]{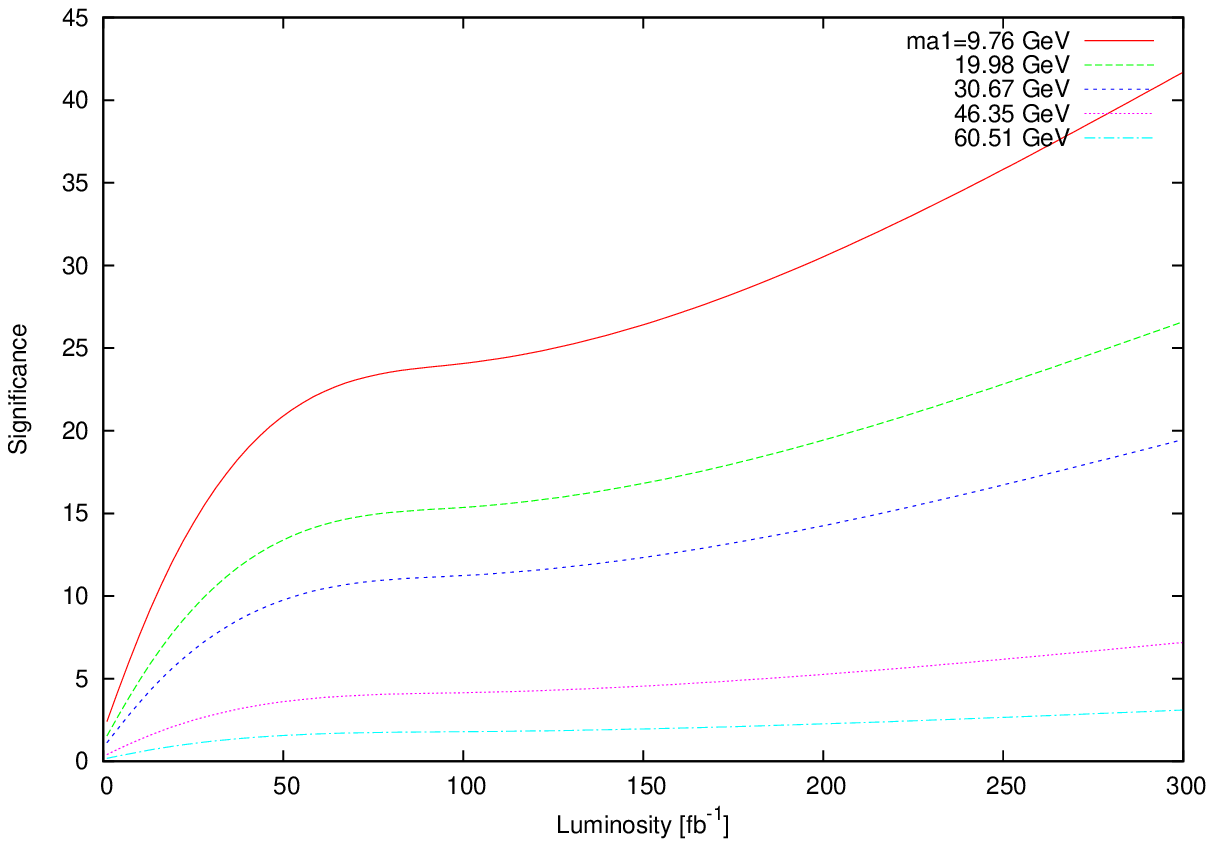}&\includegraphics[scale=0.6]{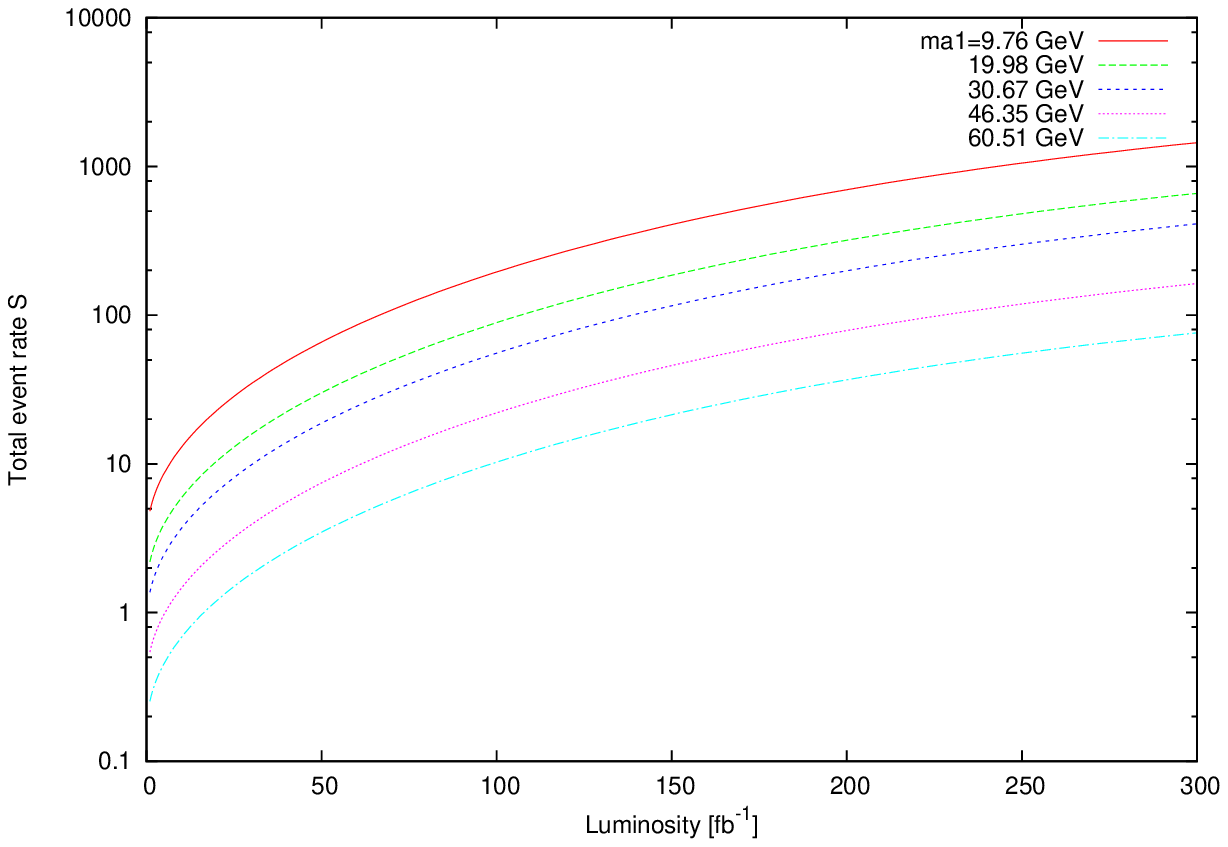}
 \end{tabular}
\label{fig:significance}
\caption{The significance $S/\sqrt B$ (left) and total event rate $S$ (right) of the $q\bar q,gg\to b\bar b a_1\to b\bar b \mu^+\mu^-$ signal as a function 
of the integrated luminosity.}
\end{figure}

\end{document}